\documentclass[12pt]{article}
\newcommand{\simB}
{\mbox{\raisebox{-0.05ex}{$\textstyle \; B$}
\raisebox{ 1.75ex}{$\textstyle \!\!\!\!\!\! \sim$  }}}

\begin{document}
\title{
\Large \bf Spherically symmetric dark energy structures in a generalized Chaplygin gas model}
\author{
{\large \bf Abiy G. Tekola}\\[0.5cm]
{Department of Physics, University of Cape Town}\\
{Private Bag, Rondebosch 7701, South Africa} }
\maketitle
\begin{abstract}
\noindent
Spherically symmetric dark energy structures are investigated in the framework of a generalized Chaplygin gas (GCG), which has an equation of state of the form $P = - A/ \rho^{\alpha}\;$.
We also study these in a modified GCG equation of state, which includes a matter term, i.e.
$P = \sigma^{2}\;\rho - A/\rho^{\alpha}$. The results of the latter are then compared with some observational data on low-surface-brightness galaxies which are supposed to be dominated by dark matter.
\end{abstract}
\newpage

\setlength{\baselineskip}{1.5\baselineskip}

\subsection*{1.$\;\;$Introduction}
The generalized Chaplygin gas is one of the promising candidates to explain the present accelerated expansion of the universe. It is a simple model unifying dark energy and dark matter based on a perfect fluid having negative pressure [1].
The equation of state of the generalized Chaplygin gas is given by
\begin{equation}
P_{ch} = - \frac{A}{\rho_{ch}^{\alpha}} \; \; ,
\end{equation}
where $A$ and $\alpha$ are positive constants and $0 \leq \alpha \leq 1$ and $P_{ch}$ and $\rho_{ch}$ are the generalized Chaplygin gas pressure and density, respectively. The energy conservation in the context of FRW cosmology yields an expression for the density $\rho$ in terms of the scale factor $a$ as
\begin{equation}
\rho_{ch} = \left( A + \frac{B}{a^{3(1+ \alpha)}} \right)^{\frac{1}{1+ \alpha}} \; \; .
\end{equation}
The corresponding expression for the original Chaplygin gas is the special case $\alpha$ = 1 of this
equation. The generalized Chaplygin gas model, like the original Chaplygin gas model, evolves first as dust and then as cosmological constant at late times. However, in its intermediate stages, it behaves as a mixture of a cosmological constant and a perfect fluid with an equation of state $P = \lambda \rho$. The parameter $\lambda$ in the generalized Chaplygin equation makes this model more flexible for comparison against observations. In this paper, we will investigate the possible existence of   spherically symmetric dark energy structures in the framework of the generalized Chaplygin Gas (GCG).
\subsection*{2.$\;\;$Structure equations and solutions}
Starting with a line element 
\begin{equation}
\displaystyle{ds^{2} = e^{\alpha}\;dt^{2} - e^{\beta}\;dr^{2} - r^{2}\; \left( d \theta^{2}\;+\;\sin^{2}\;\theta\;d \phi^{2} \right)}
\end{equation}
and choosing 
\begin{equation}
\displaystyle{e^{\beta} = \frac{1}{1\;-\;\frac{2GM}{r}} }
\end{equation}
and expressing the density in terms of the enclosed mass, the conservation equation $T_{\mu;\nu}^{\nu} = 0$ yields
\begin{equation}
P' + \frac{\alpha'}{2}\;(\rho + P) = 0
\end{equation}
and therefore
\begin{equation}
\displaystyle{  
\frac{P'}{P\;+\;\rho} }
 \; + \; 
\displaystyle{
\frac{ G  \left(   \displaystyle{ \frac{M}{r^{2}} }
 + 4 \pi r P \right) }   
     { 1\;-\; \displaystyle{\frac{2GM}{r}} }  \;\; = \;\; 0  } \; \; .
\end{equation}
Eq.(6) is a generalization of the equilibrium Euler-Poisson equation. Indeed, for small values of $P/\rho$, $r^{3}P$, and $2GM/r$, it is possible to obtain an approximation of eq.(6) as
\begin{equation}
r^{2} \; \frac{P'}{\rho} + GM = 0 \; \; .
\end{equation}
Finally, combining eq.(6) with the equation 
\begin{equation}
M' = 4 \pi\;r^{2}\;\rho \; \; ,
\end{equation}
we can get
\begin{equation}
\frac{1}{r^{2}} \; \left( r^{2} \frac{P'}{\rho} \right)' + 4 \pi G \rho = 0  \; \; .
\end{equation}
This is a useful approximate equation to find a scaling solution for any arbitrary equation of state.\\[.2cm]
In order to determine the scaling solution for the generalized Chaplygin gas model density, let us assume that the density profile, we are looking for, is 
\begin{equation}
\rho (r) = \frac{C}{(H_{0} r)^{\gamma}} \; \; .
\end{equation}
Here $H_{0}$ is the present day Hubble constant which corresponds to the critical density $\rho_{c}$, as 
$H_{0}^{2} = 8 \pi G \rho_{c}/3$, $C$ and $\gamma$
being constants to be determined. Combining equations (9) and (10) yields
\begin{equation}
\displaystyle{
\gamma \biggl[ 1 + \gamma (1 + \alpha) \biggr]}
\; 
\displaystyle{\alpha A r^{\gamma (1 + \alpha) - 2} }
\; 
\displaystyle{H_{0}^{\gamma (2 + \alpha)} }
\;
= \; 
\displaystyle{4 \pi\;Gr^{- \gamma}\;C^{2+ \alpha} } 
\; \; .
\end{equation}
Matching the powers and the coefficients of both terms on left-hand side of this equation,
the values of $\gamma$ and $C$ are found to be
\begin{equation}
\displaystyle{\gamma = \frac{2}{2 + \alpha} }
\end{equation}
and
\begin{equation}
\displaystyle{ C = \biggl[ \frac{4 \alpha (4 + 3 \alpha)}{3 (2 + \alpha)^{2}} \; A \rho_{c} 
\biggr]^{\frac{1}{2 + \alpha} } } \; \; ,
\end{equation}
respectively.
To fix $A$, let us consider a generalized Chaplygin gas dominated by the FRW
model. According to this model the square of the Hubble parameter is given as
\begin{equation}
H^{2} \; = \; \left( \frac{\dot{a}}{a} \right)^{2} \; = \;
\frac{8 \pi G}{3} \; \rho \; = \;
\frac{8 \pi G}{3} \; 
\left( A + 
\frac{B}{a^{3 (1 + \alpha)}}
\right)^
{ \frac{1}{1 + \alpha} } \; \; .
\end{equation}
The present day critical density is
\begin{equation}
\rho_{c} = (A + B)^{\frac{1}{1 + \alpha} } \; \; ,
\end{equation}
choosing $a = 1$ at present. 
The constant $B$ can also be expressed in terms of the matter fraction $\Omega$ at high
redshift ($a \ll 1$) as $B = (\Omega \rho_{c})^{1 + \alpha}$. The constant $A$ is then found to be
\begin{equation}
A \; = \; \rho_{c}^{1 + \alpha} \; \left( 1 - \Omega^{1 + \alpha} \right)
\end{equation}
and therefore,
\begin{equation}
\rho (r) \; = \; \frac{ \rho_{c} \overline{C} }
{( H_{0} r)^{2/(2 + \alpha)} } \; \; ,
\end{equation}
with
\begin{equation}
\overline{C} \; = \; \biggl[
\frac{ 4 \alpha (4 + 3 \alpha ) }{3 (2 + \alpha )^{2} } \;
\left( 1 - \Omega^{1 + \alpha} \right) 
\biggr]^{ \frac{1}{2 + \alpha} } \; \; .
\end{equation}
Based on the assumptions in order to get the approximate equation (9), this solution is subject to
the condition
\begin{equation}
\left|\frac{P(r)}{\rho (r) }\right| \; = \;
\frac{A}{ \left[ \rho (r) \right]^{1 + \alpha} } \; = \;
\left( 1 - \Omega^{1 + \alpha} \right)^{ \frac{1}{2 + \alpha} } \;
\Biggl[ \left(
\frac{ 3 (2 + \alpha)^{2} }{4 \alpha (4 + 3 \alpha )} \right)
H_{0}^{2}\;r^{2} \Biggr]^{\frac{1 + \alpha}{2 + \alpha} } \ll 1
\end{equation}
or essentially $r \ll H_{0}^{-1}$.
As can be seen from eq.(17), the spherically symmetric 
dark energy density can be as large as 10$^{3}\;\rho_{c}$, on a scale of 100 kpc. 
The metric inside the spherically symmetric Chaplygin gas is the Schwarzschild metric, while the background is a FRW metric. The spherically symmetric dark energy is embedded in the background fluid. In order to fit this spherically symmetric dark energy properly into the background, the pressure in both sides of the boundary must be equal [2-3], i.e.
\begin{equation}
P_{FRW} \; = \; P_{SCW} \; \; ,
\end{equation}
which implies that the densities in the two regions must also be equal, as we are dealing with the same Chaplygin gas equation of state both inside and outside the spherically symmetric dark energy, i.e.
\begin{equation}
\rho_{FRW} \; = \; \rho_{SCW} \; \; .
\end{equation}
The density profile inside the spherically symmetric system, with Schwarzschild metric, is
$M' = 4 \pi r^{2}\;\rho$. 
The outside fluid density at high redshift ($a \ll 1$) is then given as
\begin{equation}
\rho_{FRW} \; = \; \frac{\rho_{c} \Omega}{a^{3}} \; \; \; \mbox{for} \; \; \; 
a \ll 1 \; \; .
\end{equation}
Plugging the corresponding densities into eq.(26), we obtain
\begin{equation}
\frac{ \rho_{c} \overline{C} }
{\left( H_{0} r \right)^{\frac{2}{2 + \alpha}}} \; = \;
\frac{ \rho_{c} \Omega}{a^{3}} \; \; .
\end{equation}
Solving for the radius $r$ in terms of the scale factor $a$ and the constant $B$, we arrive at
\vspace*{-0.5cm}
\begin{equation}
r (t) \; = \; \simB\;a^{3 + \frac{3\alpha}{2} } \; \; ,
\end{equation}

\vspace{0.3cm}

with the constant 
\begin{equation}
\displaystyle{ \simB = H_{0}^{-1}\; \biggl[ \frac{\overline{C}}{\Omega} 
\biggr]^{1 + \frac{\alpha}{2} } \; \; } \; \; .
\end{equation}
\subsection*{3.$\;\;$Discussion and Results}
In their recent paper [4], Bertolami and Paramos studied a spherically symmetric dark energy
structure (they called ``dark energy''), using a polytropic equation of state of negative index.
They argued that there are conditions for such objects to form, due to density fluctuations in the background of the generalized Chaplygin gas. According to these authors, the condition that helps the initial fluctuation to grow, is that the sound velocity at the surface of the dark star must be less than the initial expansion velocity of the dark star.\\[.2cm]
According to our solution, the expansion rate is calculated to be
\begin{equation}
\dot{r} \; = \; (3 + 3\alpha/2) \; \displaystyle{ \Biggl[
\frac{\overline{C}^{2+\alpha}}{\Omega^{1+\alpha}} \Biggr]^{1/2}
a^{3 (1 + \alpha)/2 } }
\end{equation}
and the speed of sound is given by
\begin{equation}
c_{s}^{2} \;=\; \frac{\alpha A}{\rho^{1+ \alpha}} = - \alpha \; \frac{P}{\rho} 
\end{equation}
\begin{equation}
c_{s} \;=\;
\Biggl[ 
\frac{\alpha \rho_{c}^{1+ \alpha} \; \left( 1 - \Omega^{1+ \alpha} \right)}
{ \displaystyle{ \biggl[ \frac{\rho_{c} \Omega}{\alpha^{3}} \biggr]^{1+ \alpha}} } 
\Biggr]^{1/2} = \biggl[ \frac{\alpha (1 - \Omega^{1+ \alpha} )}{\Omega^{1+ \alpha}} \biggr]^{1/2} \;
a^{3 (1 + \alpha)/2 } \; \; .
\end{equation}
The ratio of the expansion rate to the speed of sound is
\begin{equation}
\frac{\dot{r}}{c_{s}} = \frac{3}{2} \; (2 + \alpha) \;
\frac{2 \sqrt{4 + 3\alpha}}{\sqrt{3} (2 + \alpha)} \; = \;
\sqrt{12 + 9\alpha} > 1 \; \; .
\end{equation}
Then it is true that the condition of Bertolami and Paramos [4] holds. However, this is not 
sufficient to conclude that such objects exist. An additional point that has to be checked, is the size of the initial density contrast. The average density is
\begin{equation}
\frac{m(r)}{ \frac{4}{3}\; \pi r^{3}} \; = \;
\frac{3 (2 + \alpha)}{4 + 3\alpha} \; \bar{\rho}(\alpha) \; \; ,
\end{equation}
which means that the density contrast is 
\begin{equation}
\displaystyle{\delta = \frac{2}{4 + 3\alpha} }\; \;.
\end{equation}
Thus this density contrast is independent of time. For particular case $\alpha$ = 1, the density contrast is about 0.28, which is much larger than the value $10^{-5}$ predicted by CMB observations.  Therefore, even if the expansion rate of the dark star is larger than the speed of sound, the density contrast is not sufficient to grow and form such structures.\\[.2cm]
Generally, a non-vanishing speed of sound is a major problem for structure formation in all unified models of dark energy and dark matter. Such systems have a characteristic scale, the sonic horizon, below which the pressure frustrates gravitational structure formation. Small perturbations of scales below this characteristic scale die off without any noticeable effect. Combining the continuity and the Euler-Poisson equations for the case of vanishing shear and rotation, subtracting the background, and finally changing the variable from $t$ to $a$, one gets [5]
\begin{equation}
\alpha^{2} \delta'' \; + \; \frac{3}{2} \alpha \delta' \; - \; \frac{3}{2} \delta (1 + \delta) \; - \;
\frac{4}{3} \;
\frac{(\alpha \delta')^{2}}{1 + \delta} \; - \;
\frac{1 + \delta}{a^{2} H^{2}} \;
\frac{\partial}{\partial x_{i}} \;
\left( \frac{c_{s}^{2}}{1 + \delta} \; 
\frac{\partial \delta}{\partial x_{i}} \right) \; = \; 0 \; \; ,
\end{equation}
where $\displaystyle{\delta = \frac{\rho - \bar{\rho}}{\bar{\rho}} }$ is the density contrast, $c_{s}$ is the speed of sound, given by
\vspace{-0.5cm}
\begin{equation}
c_{s}^{2} = \frac{dP}{d \rho} = \frac{\alpha A}{\rho^{\alpha}}
\end{equation}
for the Chaplygin gas model.\\[.2cm]
The linear solution for the perturbative density contrast of eq.(19), well discussed by Fabris {\it et al.} [29], is given as
\begin{equation}
\delta_{per} (k, \alpha) \propto\; \alpha^{-1/4} \; J_{5/14} \; (d_{s}\;k) \; \; ,
\end{equation}
where $J_{\nu}(z)$ is the Bessel function and $k$ is the comoving wave number and $d_{s}$ is the sonic horizon. This means that $\delta_{per} \sim a$ for $d_{s} k \ll 1$, and it oscillates with a decaying amplitude when $d_{s} k \gg 1$. This is because the non-zero speed of sound, and hence non-zero pressure, opposes the effect of gravity below that characteristic (sonic horizon) scale. This characteristic scale is, for the original Chaplygin gas model, given by [5]

\begin{equation}
d_{s} \; = \; \int_{0}^{a} \; 
\frac{c_{s} d \alpha}{a^{2} H} \; = \; \frac{2}{7} \;
\frac{\left( 1 - \Omega^{2} \right)^{1/2}}{\Omega^{3/2}} \;
\frac{a^{7/2}}{H_{0}} \; \; ,
\end{equation}

where
$\bar{c}_{s} = \sqrt{A}/\bar{\rho}$ and
$\Omega = \sqrt{B/(A + B)} = \sqrt{B}/ \rho_{cr} \; \;$.
The sonic horizon has a radius of about 0.18 Mpc at a redshift $z \sim 20$. This distance is much larger than the size of the largest galaxy ever observed.\\[.2cm]
It is necessary to investigate eq.(32) beyond the linear regime, in order to decide whether initial perturbations can grow to the extent that they are powerful enough to give rise to gravitational condensation or not. Generally, in the nonlinear region there is a possibility, though very small, that initial perturbations can grow unlimited. Bili\'{c}, Lindebaum, Tupper and Viollier concluded in their paper [5] that unlike the linear theory, where $\delta_{R}$ finally stops growing by acoustic horizon for any value of R, the perturbation $\delta_{R}(a)$ grows in the nonlinear regime to infinity at a finite $a$, for an initial $\delta_{R}(a_{dec})$ beyond a certain limit, like the case in the dust model.\\[.2cm]
However, they also showed in the same paper that the fraction of Chaplygin gas which condensates,
due to this infinitely large perturbation, is only 1\% [5]. This fraction is too small, in order to conclude that the Chaplygin gas favours structure formation. Being a Chaplygin gas system with a non-zero speed of sound ($c_{s}^{2} = dP/d \rho = \alpha A/\rho^{2} \neq 0$), the spherically symmetric dark energy that we are discussing is governed by the general perturbation principle of Chaplygin gas mentioned above. This implies that structure formation is impossible, as the effect of gravity is opposed by the pressure. Thus structure formation contradicts the existence of spherically symmetric dark energy structures (dark stars) in the first place.\\[.2cm]
In addition to the density contrast problem, the rotational velocity profile of the dark energy structures is not consistent with observation.
Starting from the density profile determined in eq.(17), the enclosed mass can be calculated as
\begin{equation}
m (r) \; = \; \frac{4 \pi (2 + \alpha)}{4 + 3 \alpha} \; \rho_{c} \; \overline{C}\;H_{0}^{- \gamma}\;
\displaystyle{r^{\frac{4+3 \alpha}{2+ \alpha}} } \; \; .
\end{equation} 
The resulting rotational speed inside the spherically symmetrical dark energy (it could be a halo of a galaxy or even a certain spherical patch inside a galaxy), is then given by
\begin{equation}
V_{c} \; = \; \Biggl[ \frac{3}{2}\; \frac{2+ \alpha}{4+3 \alpha}\; \overline{C} \Biggr]^{1/2}
\left( H_{0} r \right)^{ \frac{1+ \alpha}{2+ \alpha} } \; \; .
\end{equation}
For the special case $\alpha = 1$, the rotational velocity is proportional to $r^{2/3}$. This relation completely contradicts the flat rotational curve observed for galaxies.\\[.2cm]
However, it is possible to fix this problem by modifying the pressure and the density relation [6]. This can be done by combining a Chaplygin gas pressure with a pressure that depends on density linearly, i.e.
\begin{equation}
P \; = \; \sigma^{2} \; \rho \; - \; \frac{A}{\rho^{\alpha}} \; \; .
\end{equation}
According to this modified pressure, the body would have a flat rotational curve at high density, because at such densities, only the first term is dominant. The problem arises when the density is too low. In that case, the second term, the generalized Chaplygin gas term, is dominant and we are faced with the same non-flat rotational curve problem. It is, however, possible to determine the maximum distance or crossover point to which the first term dominates. This point is where the pressures are equal.
$P = \sigma^{2} \;\rho$ and that of the generalized Chaplygin gas  $P = - A/\rho^{\alpha}$ with
\begin{equation}
\displaystyle{\rho (r) = \frac{\rho_{c} \overline{C}}{\left(H_{0} r \right)^{2/(2+ \alpha)}} }
\end{equation}
and
\vspace{-0.5cm}
\begin{equation}
\displaystyle{ \overline{C} \; = \; \Biggl[
\frac{4 \alpha (4 + 3 \alpha)}{3 (2 + \alpha)^{2}} \; \left( 1 - \Omega^{1 + \alpha} \right)
\Biggr]^{\frac{1}{2+ \alpha}} } 
\end{equation}
are equal. The density and the rotational velocity corresponding to the equation of state,
$P = \sigma^{2}\; \rho$, of the first part in the right-hand side of eq.(29) are given respectively as [7]
\begin{equation}
\rho = \frac{\sigma^{2}}{2 \pi Gr^{2}} \; \; \; \mbox{and} \; \; \; V_{c} = \sigma\;\sqrt{2} \; \; .
\end{equation}
The constant $\sigma$ can be fixed by comparing the second equation with the value of the circular velocity from a real rotational curve.\\[.2cm]
Plugging in the corresponding values of the densities and the value of the constants $A$ and the constant $\sigma$ into eq.(24) and finally solving for the crossover radius $r_{c}$, we obtain
\vspace{-0.5cm}
\begin{equation}
r_{c} \; = \; H_{0}^{-1}\; \Biggl[
\frac{V_{c}^{4} \; \overline{C^{\alpha}}}{3 \left( 1 - \Omega^{1+ \alpha} \right)}
\Biggr]^{\frac{2+ \alpha}{4(1+ \alpha)}} \; \; ,
\end{equation}
where $V_{c}$ is the rotational velocity, which varies from galaxy to galaxy.\\[.2cm]
In order to get some numerical insight into the the comparison between the crossover radius and the actual radius of some LSB (low-surface-brightness) and HSB (high-surface-brightness) galaxies, the crossover radius is calculated using the above formula for a Chaplygin gas with $\alpha = 1$ for both LSB and HSB galaxies with their observed rotational velocity. A value of 71 km/s/Mpc and 0.28 are taken for the Hubble constant, $H_{0}$ and $\Omega$, respectively.\\[.2cm]
The real radius, the rotational velocity and the corresponding calculated value of the crossover radius for both LSB and HSB galaxies are given in Table 1 as de Block and McGaugh mentioned in their paper [8].

\newpage

\vspace*{-3.0cm}

\begin{center}
\begin{tabular}{lrrrcrrr}
{\bf Name} & {\bf $V_{\rm max}$} & {\bf $R_{\rm max}$(\small kpc)} & {\bf $R_{\rm cross}$(\small Mpc)} & \hspace{1cm} &
{\bf Name} & {\bf $V_{\rm max}$} & {\bf $R_{\rm max}$(\small kpc)}\\ \hline
           &               &         &       & &        &    &    \\
   F561-1 & 52  & 10.1 & 3.59  & & DDO154 & 48 & 7.6\\[.1cm] 
   F563-1 & 111 & 17.7 & 11.18 & & DDO168 & 55 & 3.4\\[.1cm] 
   F563V1 & 30  & 7.4 & 1.57  & & DDO170 & 66 & 9.6\\[.1cm] 
   F563V2 & 111 & 9.2 & 11.18 & & N55 & 87 & 10.2\\[.1cm] 
   F564V4 & 40  & $-$2 & 2.42  & & N257 & 108 & 9.9\\[.1cm] 
   F565V2 & 51 & 8.4 & 3.48 & & N300 & 97 & 10.6\\[.1cm] 
   F567-2 & 64  & 11.3 & 4.90  & & N801 & 222 & 58.7\\[.1cm] 
   F568-1 & 119 & 14.9 & 12.42 & & N1003 & 115 & 31.3\\[.1cm] 
   F568-3 & 120  & 16.5 & 12.57  & & N1530 & 79 & 8.3\\[.1cm] 
   F568V1 & 124 & 19 & 13.21 & & N2403 & 136 & 19.5\\[.1cm] 
   F571-8 & 133  & 15.6 & 14.67  & & N2841 & 323 & 81.1\\[.1cm] 
   F571V1 & 73 & 14.6 & 5.97 & & N2903 & 201 & 24.2\\[.1cm] 
   F571V2 & 45  & 3.7 & 2.89  & & N2998 & 214 & 46.6\\[.1cm] 
   F574-1 & 100 & 15.4 & 9.56 & & N3109 & 67 & 8.2\\[.1cm] 
   F574-2 & 40  & 10.7 & 2.42  & & N3198 & 157 & 29.9\\[.1cm] 
   F577V1 & 30 & 8.9 & 1.57 & & N5033 & 222 & 35.4\\[.1cm] 
   F579V1 & 100 & 17.3 & 9.56 & & N5533 & 273 & 74.4\\[.1cm] 
   F583-1 & 85  & 14.6 & 7.49  & & N5585 & 92 & 9.6\\[.1cm] 
   F583-4 & 67 & 10 & 5.25 & & N6503 & 121 & 22.2\\[.1cm] 
   U0128  & 131 & 42.3 & 14.34 & & N6674 & 266 & 64.5\\[.1cm] 
   U1230  & 102  & 34.7 & 9.85  & & N7331 & 241 & 36.7\\[.1cm] 
   U5005 & 99 & 27.8 & 9.42 & & U2259 & 90 & 7.6\\[.1cm] 
   U5750 & 75 & 21.8 & 6.21 & & U2885 & 298 & 72.5\\[.1cm] 
   U5999 & 155 & 15.3 & 18.46 & &  &  & \\
\end{tabular}    
\end{center}

\vspace{0.5cm}

\setlength{\baselineskip}{1\baselineskip}

TABLE 1: The table on the left shows the rotational speed, the radius and the corresponding crossover radius for LSB galaxies. Similarly, the table on the right shows the rotational speed, the radius and the corresponding crossover radius for HSB galaxies [8].\\[.2cm]

\newpage

As can be seen from Table 1, the real radius of both the LSB and HSB galaxies is of order kpc. However, the corresponding crossover radius for each galaxy is of order Mpc. Thus this modified equation of state works as far as a couple of hundreds Mpc. There is a huge difference between the real radius and the crossover radius. This means that the point at which the modified equation of state fails to hold, is out of the physical extent of these galaxies and hence the modified Chaplygin equation of state can be applied, and it can adequately explain the flat rotational curves. Therefore this equation could be a good alternative to the Chaplygin gas equation, as it could possibly alleviate the rotational curve problem which the Chaplygin gas causes at low densities for spherically symmetric dark energy structure.\\[.2cm]
It is clear that $\sigma$ must be a constant that is not characteristic of a particular galaxy. However, the rotational velocity $V_{c}$ varies from galaxy to galaxy and since $\sigma$ depends on the rotational velocity as $\sigma = V_{c}/\sqrt{2}$, $\sigma$ then differs from galaxy to galaxy. The main problem is generalizing such models as one theory which governs different galaxies.\\
\subsection*{4.$\;\;$Conclusions}
The unrealistic rotational velocity, the very less value of density contrast, together with the small probability (1\%) of getting an initial density fluctuation which can result in the formation of large-scale structure exhibited by the spherically symmetrical dark energy structure, leads to the inevitable conclusion that such objects cannot exist under the context of a generalized Chaplygin Gas (GCG) model.\\[.2cm]
We then modified the GCG equation of state to $P = \sigma^{2} \; \rho - A/\rho^{\alpha}$. At high density the first term dominates and the resulting rotational curve is consistent with the flat rotational curves observed for galaxies, but at low density the rotational curve is unrealistic. Two types of galaxies (LSB and HSB) are considered to compare the real radius of these sample galaxies with the radius to which the modified equation of state is working for a special case $\alpha = 1$. Fortunately, the point from which this unrealistic rotational curve starts to dominate is completely out of the physical size of the sample galaxies considered. Therefore, this modified equation of state looks capable of explaining the spherically symmetrical dark energy rotational curve. The problem is as the constant $\sigma$ differs from galaxy to galaxy, there is no general equation, which holds for all galaxies.\\   
   
\noindent
{\large \bf Acknowledgement}\\
\noindent
I would like to thank G.B. Tupper and R.D. Viollier for valuable discussions and comments. This research is supported by the National Astrophysics and Space Science Programme (NASSP) bursary.\\

\newpage

\end{document}